\documentclass[12pt]{iopart}
\usepackage{graphicx, multirow, braket, ulem}

\begin{document}
\interfootnotelinepenalty=100000

\title[High speed flux sampling for tunable superconducting qubits]{
High speed flux sampling for tunable superconducting qubits with an embedded cryogenic transducer}

\author{B. Foxen,$^{1}$ J.Y. Mutus,$^{2}$ E. Lucero,$^{2}$ E. Jeffrey,$^{2}$ D. Sank,$^{2}$ R. Barends,$^{2}$ K. Arya,$^{2}$  B. Burkett,$^{2}$ Yu Chen,$^{2}$ Zijun Chen,$^{2}$ B. Chiaro,$^{1}$ A. Dunsworth,$^{1}$ A. Fowler,$^{2}$ C. Gidney,$^{2}$ M. Giustina,$^{2}$ R. Graff,$^{2}$ T. Huang,$^{2}$ J. Kelly,$^{2}$ P. Klimov,$^{2}$ A. Megrant,$^{2}$ O. Naaman,$^{2}$ M. Neeley,$^{2}$ C. Neill,$^{2}$ C. Quintana,$^{2}$ P. Roushan,$^{2}$  A. Vainsencher,$^{2}$ J. Wenner,$^{1}$ T.C. White,$^{2}$ 
and John M. Martinis$^{1,2,b)}$}

\address{$^1$ Department of Physics, University of California, Santa Barbara, CA 93106-9530}
\address{$^2$ Google, Santa Barbara, CA 93117}
\ead{$^b$ martinis@physics.ucsb.edu}
\vspace{10pt}

\begin{abstract}

We develop a high speed on-chip flux measurement using a capacitively shunted SQUID as an embedded cryogenic transducer and apply this technique to the qualification of a near-term scalable printed circuit board (PCB) package for frequency tunable superconducting qubits. The transducer is a flux tunable LC resonator where applied flux changes the resonant frequency. We apply a microwave tone to probe this frequency and use a time-domain homodyne measurement to extract the reflected phase as a function of flux applied to the SQUID. The transducer response bandwidth is 2.6\,GHz with a maximum gain of $\rm 1200^\circ/\Phi_0$ allowing us to study the settling amplitude to better than 0.1\%. We use this technique to characterize on-chip bias line routing and a variety of PCB based packages and demonstrate that step response settling can vary by orders of magnitude in both settling time and amplitude depending on if normal or superconducting materials are used. By plating copper PCBs in aluminum we measure a step response consistent with the packaging used for existing high-fidelity qubits. 

\end{abstract}

\vspace{2pc}

%
%
%
%
%

\section{Introduction}

Superconducting qubits have risen to a position of prominence in the field of quantum computation as they have demonstrated a system-level performance near the fault-tolerant threshold \cite{Barends2014}.  The demonstration in ref.\,\cite{Barends2014} was an important milestone, and further performance gains above and beyond the error correction threshold will exponentially lower the required resources \cite{PhysRevA.86.032324}.  While qubit decoherence remains the largest source of system infidelity, control error in two-qubit gates is a close second \cite{jimmythesis}. Frequency tunable superconducting qubits, which use a superconducting quantum interference device (SQUID) loop as a magnetic-flux-tunable non-linear inductance, perform entangling operations by bringing qubit frequencies near resonance with shaped flux pulses.  The highest fidelity two-qubit entangling gates have used a fast-adiabatic approach which is theoretically capable of realizing a CZ gate with error less than $10^{-4}$ \cite{PhysRevA.90.022307}.  Experimental implementations of this gate have thus far achieved error rates of $5.4\cdot 10^{-3}$ per gate, where two qubit randomized benchmarking and purity measurements in that system indicate qubit decoherence accounts for about 60\% of the gate error while control error is estimated to account for most of the remaining 40\% \cite{jimmythesis}. Characterizing the control error is therefore a priority for quantum computation with superconducting qubits.

Flux control pulses for qubits are often calibrated \textit{in situ} to account for line-specific impedance varaitions due to connectors, wire bonds, and other imperfections that alter these flux pulses before they reach the qubits \cite{PhysRevLett.112.240504, Neill195}. However, the time resolution of these measurements is limited by the width of microwave spectroscopy pulses used to probe the qubit frequency, which are typically 5-20\,ns. In this paper we demonstrate a time-domain, qubit-free, GHz bandwidth measurement for flux pulses using a capacitively shunted SQUID as a flux-tunable resonator to realize a cryogenic magnetic flux to microwave phase transducer. In this transducer, flux pulses shift the resonator frequency which in turn modulates the phase of a microwave tone reflecting off of the resonator. Similar circuits have been operated as quantum limited amplifiers optimized for flux sensitivity at the cost of bandwidth \cite{PhysRevB.83.134501}; we have instead optimized for a large bandwidth at the expense of flux sensitivity by operating without amplification and using an impedance transformer to reduce the resonator's quality factor. In contrast to experiments that measure flux settling with qubits, this measurement may be performed with just a single microwave source, a two channel oscilloscope, and an arbitrary waveform generator (AWG). We achieve a DC-2.6\,GHz flux waveform measurement bandwidth which may be used to measure pulse settling to better than 0.1\% of a step with an absolute flux sensitivity of $\rm 0.2\,m\Phi_0$.
	
In section 2 of this paper, we provide a simple analytic model of the tunable resonator device. In section 3, we describe the measurement setup and DC calibration of the tunable resonator as a flux-to-phase transducer. In section 4, we verify the performance of the tunable resonator by measuring the settling response in a package we can compare with previous qubit results. Finally, in section 5, we apply this measurement technique to characterize a variety of device packages yielding a near-term scalable PCB-based package and gaining some insight into how to further improve the performance of our qubit flux bias lines. Notably, in \ref{appendix:microwave_reflections} we bound the error introduced by microwave reflections and offer an alternate measurement technique that uses the tunable resonator to characterize microwave reflections in a cryogenic environment with a sensitivity better than -30\,dB.

\section{Theory of Operation}

\begin{figure}[h]
\begin{center}
\includegraphics[width=1.0\textwidth]{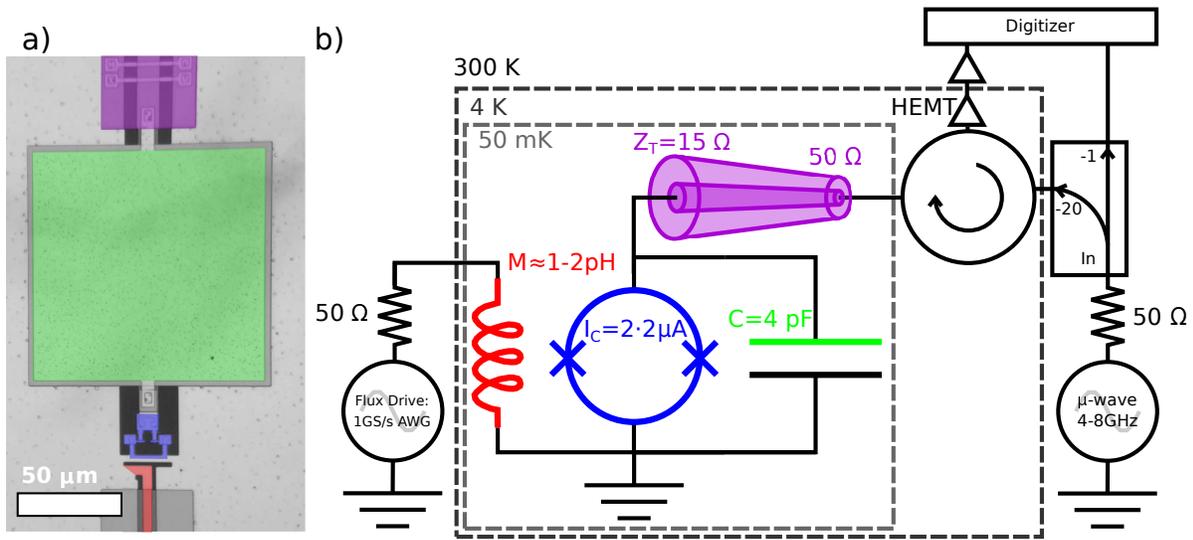}
\caption{Device and circuit diagram. a) Micrograph image of the tunable resonator which consists of a 2-junction SQUID (highlighted blue) shunted by a parallel plate capacitor (green).  Microwaves are reflected off of the tunable resonator through a galvanicly coupled $50\,\Omega$ to $15\,\Omega$ impedance transformer (magenta). Magnetic flux is applied to the SQUID through a mutual inductance with the adjacent flux bias line (red). b) Schematic diagram of the turnable resonator and measurement system.  Flux pulses are generated by a FPGA-based arbitrary waveform generator on the left. These flux pulses change the frequency of the tunable resonator which modulates the phase of the reflected probe tone. The probe tone is supplied by a variable frequency microwave source; it reflects off of the tunable resonator is directed via a cryogenic microwave circulator through amplifier stages and towards a digitizer. \label{fig:circuit_and_device}}
\end{center}
\end{figure}

The circuit used in this experiment consists of an on-chip impedance transformer terminated by a lumped element capacitor in parallel with a two-junction SQUID, as shown in figure \ref{fig:circuit_and_device}a. This circuit and its fabrication are the same as the Impedance Matched Parametric Amplifier (IMPA) in reference \cite{doi:10.1063/1.4886408}. We choose a junction critical current of 2\,$\rm \mu A$ ($\rm I_c= 2\cdot 2\,\mu$A) with a 4\,pF parallel plate capacitor giving a maximum resonance frequency of $\rm 8.7\,GHz$ and making the device compatible with the same microwave components used for typical 4-8\,GHz qubit circuits. The impedance transformer maximizes the flux measurement bandwidth by reducing the resonator's quality factor. Since $\rm Q = \omega_{r}/\Delta\omega_{r}\,=\,Z_{0}\omega_{r}C$, the resonator bandwidth $\Delta\omega_{r} = 1/(Z_0 C)$ is independent of the applied flux. With $\rm Z_0 = 15\,\Omega$ the tunable resonator is able to operate over a 2.6\,GHz bandwidth which is more than sufficient to characterize flux control pulses that typically have a bandwidth of several hundred MHz.

The flux bias line, highlighted in red in figure \ref{fig:circuit_and_device}a, has the same geometry as used in qubit circuits and shares a mutual inductance of 1-2\,pH with the SQUID loop, highlighted blue, which converts current in the bias line to a magnetic flux through the SQUID loop. This flux $\rm \Phi$ changes the inductance of the tunable resonator as given by the Josephson inductance,

\begin{equation}\label{eq1}
{\rm L_{j}(\Phi) = \frac{\Phi_{0}}{2 \pi I_{c} |\cos(\frac{\pi \Phi}{\Phi_{0}})|} = \frac{\Phi_{0}}{2 \pi I_{c}^*(\Phi)}}
\end{equation}
where $\rm I_{c}$ is twice the critical current of each of the two, nominally identical, Josephson junctions that compose the SQUID, $\Phi_{0} = h/2e$ is the magnetic flux quantum, and $\rm I_c^*(\Phi)$ is the SQUID's flux-dependent effective critical current.

The SQUID's effective critical current goes to zero as the applied flux approaches $\rm \frac{1}{2}\,\Phi_0$. As such, there will always be a flux where the current from the microwave tone will break the small amplitude limit invalidating equation \ref{eq1}\textemdash this sets the upper limit for the tunable resonator's flux operating range. Simple analysis of the circuit in figure \ref{fig:circuit_and_device}b can be used to compute the frequency dependent resonator impedance, as a function of the flux through the SQUID loop.

\begin{equation}\label{eq2}
{\rm Z_{r}(\omega, \Phi) = (L_{j}(\Phi) || C_{s}) = \frac{\dot{\imath} \omega L_{j}(\Phi)}{1 - \omega^2 L_{j}(\Phi) C} =  \frac{\dot{\imath} \omega L_{j}(\Phi)}{1 - \omega^2 /\omega_{r}(\Phi)^2}}
\end{equation}

From here it is straightforward to compute the complex coefficient of reflection for the tunable resonator, $\rm \Gamma_{r} = (Z_r - Z_0)/(Z_r + Z_0)$. Since $\rm Z_{r}$ is purely imaginary the magnitude of $\rm \Gamma_{r}$ is 1 and all of the flux dependence is contained in the reflection angle $\rm \angle \Gamma_{r}$ given by

\begin{eqnarray}
\rm \angle \Gamma_{r} &= \rm \arctan\left({\frac{Im(\Gamma_r)}{Re(\Gamma_r)}}\right) = \arctan\left( \frac{-2Im(Z_r)Z_0}{Z_r^2 + Z_0^2}\right)\nonumber\\ & = \rm \arctan \left(\frac{2 Z_{0}}{\omega L_{j}(\Phi, I_c)} \frac{((1-\omega^2 L_{j}(\Phi, I_c) C)^2)^2 - Z_{0}^2)}{(1-\omega^2 L_{j}(\Phi, I_c) C)}\right). \label{eq3} 
\end{eqnarray}

\noindent Where $\rm \omega$ is the frequency of the incoming probe tone, and $\rm Z_0$, $\rm I_c$, and C are our circuit design parameters. Equation \ref{eq3} summarizes the operation of the tunable resonator.  If a single frequency probe tone $\rm \omega$ is reflected off of the tunable resonator, then it functions as a flux-to-phase transducer, translating the flux waveform applied to the SQUID to the microwave frequency domain as a phase modulation of the reflected probe tone. 

The transducer's inherent frequency up-conversion enhances its functionality as an on-chip flux transducer. For the purpose of calibrating a qubit flux bias line, we would like to measure the transfer function of the cabling, connectors, wirebonds, \textit{etc.} between the generator and a cryogenic SQUID. This is complicated by the fact that both our generation and digitization instruments are at room temperature; we can only directly measure the round-trip transfer function. However, since the SQUID upconverts the signal bandwidth, this means we capture the baseband transfer function from room temperature to the SQUID and the microwave frequency transfer function on the return path. Since dispersive effects are larger for our incoming baseband flux waveform than the outgoing phase modulated microwave signal, we are more sensitive to these effects in the desired portion of the signal path.

Finally, since the transducer's flux-to-phase response is nonlinear, we are able to operate at different flux biases to take advantage of the transducer's gain and characterize our system. By operating at a flux bias where the transducer is is maximally sensitive to flux (e.g. when the resonator and probe frequencies are equal), we are able to characterize irregularities in the flux drive line with a high precision. Alternatively, if we operate where the transducer is minimally sensitive to flux (e.g. with the resonator detuned from the probe frequency), then the transducer may be used to characterize systematic phase errors due to reflections in the microwave cabling. For a discussion of these errors, see \ref{appendix:microwave_reflections}.

\section{Measurement Setup and Transducer Calibration}
A simplified measurement schematic is shown in figure \ref{fig:circuit_and_device}b. The tunable resonator is cooled to 50\,mK in an adiabatic demagnetization refrigerator (ADR). Flux waveforms are produced at room temperature by a FPGA controlled 1 GS/s DAC that together form an arbitrary waveform generator (AWG). We place a 220\,MHz Gaussian low pass filter on the output of the AWG to prevent ringing, and stagger attenuators from room temperature down to the 4\,K stage of the ADR to reduce noise and reflections while bringing the full scale AWG output level down to 1.5-2.0\,$\Phi_{0}$ on-chip, thereby using the AWG's full dynamic range.

A variable frequency microwave source (Hittite HMC-T2220) provides the probe tone, which we split at room temperature with a directional coupler; the transmitted port is used as the local oscillator reference for our phase measurement and the coupled (-20\,dB) port is sent into the cryostat to reflect off of the tunable resonator. After the probe tone reflects off of the tunable resonator, a cryogenic microwave circulator directs the reflection through a HEMT amplifier at 4\,K and towards the room temperature amplification and digitizer. The probe frequency may be chosen arbitrarily, but practically, it should be in the pass band of the HEMT amplifier (4-8\,GHz). It is best if the tunable resonator frequency can sweep through the probe frequency to maximize the phase shift and resulting flux sensitivity\footnote{Since neither our AWG or microwave source accept external triggers, we are also required to chose a flux waveform repetition period that is a multiple of the probe tone. This way the AWG may be used to trigger the oscilloscope and the phase-locked probe tone will remain synchronized throughout the experiment. \label{triggering}}. With a maximum resonator frequency of 8.7\,GHz, we found using a probe frequency of 6.4\,GHz to be an optimal operating point. 

We measure the microwave signal with an effective homodyne receiver that is phase-locked to the AWG and microwave source through a shared 10\,MHz clock. To reconstruct the baseband phase waveform, we digitize both the local oscillator reference and the reflection with a 40\,GS/s oscilloscope (Keysight DSA90804A) and digitally demodulate the signal. Alternatively, in section 5 we use an IQ mixer to demodulate the reflection so that we may digitize at baseband frequencies and accommodate the oscilloscope's single-trace memory limit while sampling at longer times. In \ref{appendix:qubit_detune_spectroscopy} we compare digital and hardware demodulation of this tunable resonator measurement to qubit-based settling measurements. Hardware demodulation is quite precise (just a little less accurate), allows for sparse sampling at long times, and requires a significantly less expensive oscilloscope since only several hundred MHz (rather than 4-8\,GHz) of bandwidth are needed. Digital demodulation offer a superior performance with respect to flatness and settling due to the oscilloscope's superior AC \textit{vs.} DC response. 

\begin{figure}[h]
\begin{center}
\includegraphics[width=1.0\textwidth]{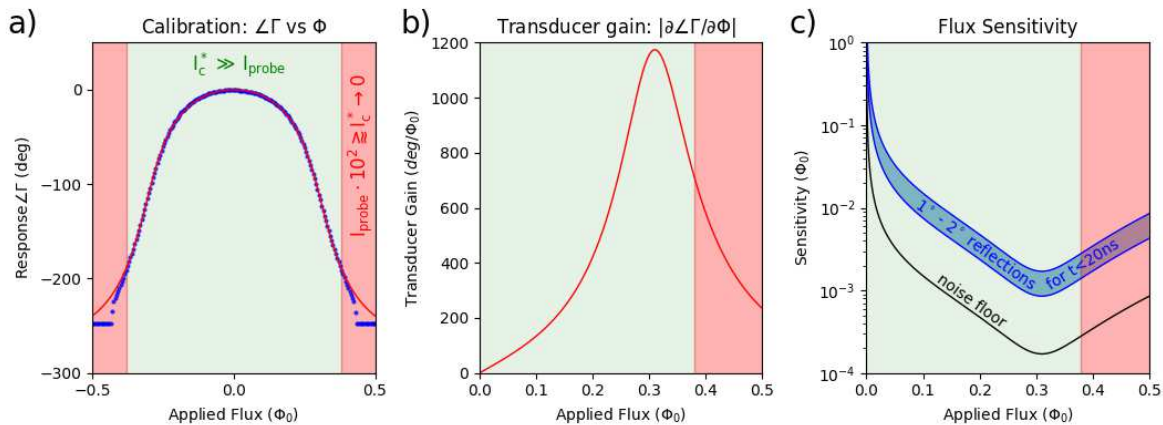}
\caption{Transducer calibration, gain, and flux sensitivity. a) Reflection angle ($\angle \Gamma$) vs. applied DC flux; this curve serves as the transducer's flux-to-phase calibration curve. For a probe frequency of $\rm 6.4\,GHz$, we fit this data using equation \ref{eq3}, and find good agreement with the designed parameters for SQUID critical current $\rm I_{c} = 1.8\,\mu$A ($\rm 2\,\mu$A designed), transformer impedance, $\rm Z_{0,fit} = 14.8\,\Omega$ $\rm (15\,\Omega)$ and shunt capacitance, $\rm C_{fit} = 3.8\,pF$ $\rm (4\,pF)$. As the flux approaches $\frac{1}{2}\,\Phi_{0}$ the effective critical current of the SQUID is reduced to zero; the fit diverges from the data at $|0.38\,\Phi_{0}|$ because the probe tone drive power is no longer in the linear limit relative to the SQUID critical current. b) Transducer gain as a function of applied flux showing a maximum at $\rm 0.31\,\Phi_{0}$. c) Flux sensitivity vs. applied flux. Our measurement sensitivity in terms of flux is limited by our phase measurement noise of about $0.25^\circ$. For short times (less than twice the propagation time from the resonator to the digitizer, approximately $\rm \,20\,ns$ in our case), the blue region represents an additional systematic error of $1^\circ$-$2^\circ$ due to transient reflection of the probe tone discussed in \ref{appendix:microwave_reflections}. \label{fig:microwave_reflections}}
\end{center}
\end{figure}

As shown in figure \ref{fig:microwave_reflections}a, we calibrate the tunable resonator as a flux-to-phase transducer by applying DC fluxes to the SQUID and measuring the reflected phase $\rm \angle \Gamma$. For a probe frequency of $\rm 6.4\,GHz$, we use equation \ref{eq3} to fit $\rm \angle\Gamma$ as a function of the applied flux and find fit parameters within 10\% of the designed parameters for junction critical current, $\rm I_c=1.8\,\mu A\,(designed = 2\,\mu$A), transformer impedance, $\rm Z_0=14.8\,\Omega\,(15\Omega$), and capacitance, $\rm C=3.8\,pF\,(4\,pF)$. While a phase shift of $360^\circ$ is typically expected when sweeping through a resonance, the phase shift we measure is reduced due to the finite (and large) resonator bandwidth relative to its flux-tunable frequency range. In order to keep the transducer's flux-to-phase relationship single-valued, we limit the operating range to a single $\frac{1}{2} \Phi_0$ bias range (e.g. $0-\frac{1}{2}\Phi_0$), with the actual upper operating limit set by the point where the probe current begins driving the SQUID non-linearly. In figure \ref{fig:microwave_reflections}a, this non-linear behavior may be observed as the point where the fit deviates from the measured phase at a flux of $|0.38\,\Phi_0|$. In this case, the probe power was chosen to be as large as possible while still allowing for operation at $0.31\,\Phi_0$ where the transducer gain is at a maximum as shown in figure \ref{fig:microwave_reflections}b. 

The accuracy and noise of our phase measurement, divided by the transducer gain in \ref{fig:microwave_reflections}b, determine our flux accuracy and sensitivity. Immediately after a large change in flux, reflections in the microwave side of the measurement dominate our error as described in \ref{appendix:microwave_reflections}. We observe a -30\,dB to -36\,dB mircrowave reflections, typical of SMA connections and other microwave components, that result in time dependent phase errors of $1^\circ$ to $2^\circ$. Since second-reflections from the same components generate -60\,dB reflections ($\rm 0.05^\circ$ errors), their effect dies off in twice the signal propagation time from the tunable resonator to the digitizer (about 20\,ns). The noise in our phase measurement sets the limit of our flux sensitivity. When digitally demodulating the settling response, the phase measurement noise is limited by the $\rm 20\,ps_{pk-pk}$ synchronization jitter between our microwave source and AWG. By averaging 50k oscilloscope traces we are able to reduce this jitter to $\rm 0.25^\circ$ for a 6.4\,GHz probe frequency. Further averaging would reduce this noise, but time-averaging samples often becomes more economical when looking at long times. In figure \ref{fig:microwave_reflections}c, we plot the flux sensitivity due to a $\rm 0.25^\circ$ phase measurement error which is better than $\rm 0.2\,m\Phi_0$ for a small range of applied fluxes around $\rm 0.31\,\Phi_0$. We also plot the systematic error in flux due to $\rm 1^\circ$ to $\rm 2^\circ$ microwave reflections at short times.

\section{Step Response}

\begin{figure}[h]
\begin{center}
\includegraphics[width=0.8\textwidth]{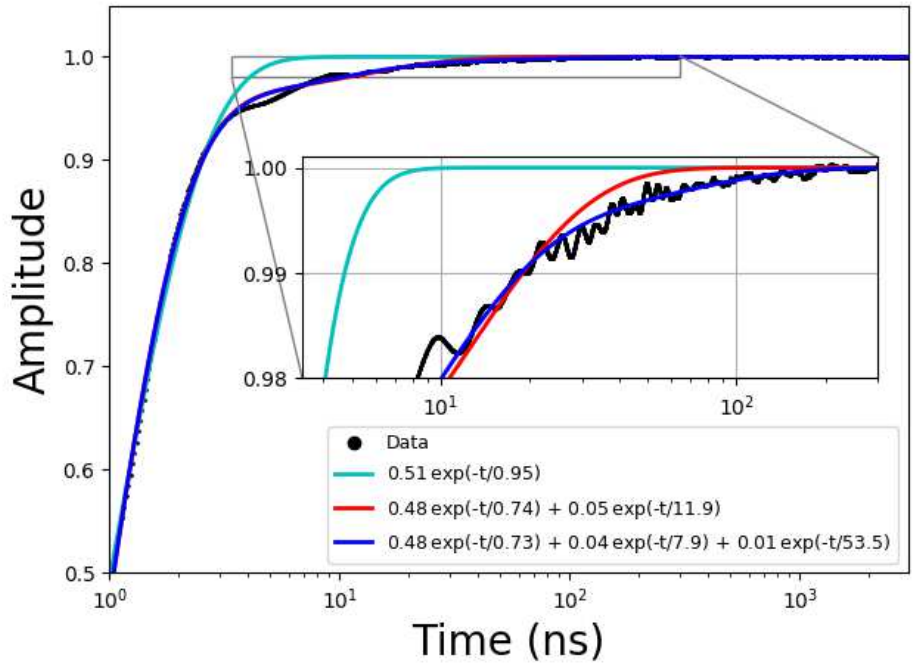}
\caption{Fitting the measured step response using equation \ref{eq5} with 1, 2, or 3 exponential decay rates. Previous measurements with the same AWG and similar cabling, packaging, and on-chip layouts found time-constants of 5\,ns and 100\,ns attributed to reflections and a L/R time constant from a bias tee.  In this measurement, we find the response to be well fit by three time constants of 0.73\,ns, 7.9\,ns and 53.5\,ns.  \label{fig:good_data}}
\end{center}
\end{figure}

We measure the settling response of the flux bias line, by using the AWG to generate a rising edge that has been digitally low pass filtered with a 220\,MHz Gaussian profile to prevent ringing. We choose to end the step at a flux of $\rm 0.31\,\Phi_0$, where the transducer is most sensitive to flux, and start the step at $\rm 0.08\,\Phi_0$ to maximize the amplitude of the step and increase the fractional sensitivity of the measurement. In figure  \ref{fig:good_data} we plot the measured flux waveform settling, A(t) (normalized to a step amplitude of 1), and fit it heuristically with three exponential time constants:
\begin{equation}\label{eq5}
{\rm A(t) = 1 - (\alpha_0 e^{-t/\tau_0} + \alpha_1 e^{-t/\tau_1} + \alpha_2 e^{-t/\tau_2})
}
\end{equation}
We find that the settling response is fit well for settling amplitudes of 0.48, 0.04, and 0.01 for time constants of 0.73\,ns, 7.9\,ns, and 53.5\,ns respectively. Previous qubit based measurements with the same model AWG found the settling to be well fit by two time constants of 5\,ns and 100\,ns attributed to reflections and an L/R time constant associated with a bias tee \cite{Barends2014}.  In the present measurement, we have removed the bias tee, so the longest time constant in our system has been reduced as expected. Additionally, since we are also able to measure at much shorter times than qubit experiments allow, we are able to fit the rising edge with a time constant of 0.73\,ns.  Since the settling is well-fit by exponential decays, it is likely that L/R time constants, rather than skin effect losses that would result in a $\rm 1/\sqrt[]{t}$ dependence, are the dominant source of pulse distortion in our system. This step response measurement's agreement with previous qubit-based results confirms that the tunable resonator technique is sufficient for characterizing and evaluating the transfer function of high performance qubit flux bias lines.

While this settling performance is imperfect, the settling amplitudes and times are sufficiently good that we may pre-distort control waveforms with relatively small correction amplitudes. It is theoretically possible to compensate for any transfer function, but larger settling amplitudes require larger compensation amplitudes which effectively reduce the signal to noise ratio of the AWG. Additionally, if the settling times ($\rm \tau's$) require compensation at times longer than our desired gate times (20-40\,ns) then it is difficult to design discretized, settling-compensated flux pulses that may be loaded into FPGA memory and sequenced arbitrarily to perform gate operations. For both of these reasons it is important to understand the source of these settling times.

\section{Applications: Package Design and Device Layout}
Most qubit and resonator measurements from our group have been performed using a machined aluminum package. This package consists of two pieces of 6061 aluminum (a base and a lid) and was designed to provide a continuous ground plane and shielding from external magnetic fields. Electrical signal connections are made by drilling holes from the perimeter of the base towards an interior chip cavity and inserting narrow copper printed circuit board (PCB) strips with dimensions appropriate to maintain a 50\,$\Omega$ impedance. On the exterior of the package, these PCB traces are soldered to the center pin of normal-metal panel-mount SMA connectors attached directly to the aluminum package. On the interior of the package, these PCB traces are wire bonded to the device under test (DUT) and ground plane wirebonds connect from the DUT to the bulk of the aluminum package. The key point for later is that in this package the signal tranmission lines have a normal metal (copper) center trace and a superconducting aluminum ground plane.

Despite this package's solid performance, its major disadvantage is that the physical machining necessary to incorporate additional signal lines is not scalable beyond a couple dozen qubits. In fact, package design and scaling up qubit systems is an active area of research for many superconducting qubit groups \cite{Rosenberg2017, PhysRevApplied.6.044010, 2058-9565-3-2-024007}. PCB-based packages, able to implement additional signal lines in lithography rather than with physical machining, offer a path forward, but we have found that traditional copper PCBs perform poorly with considerable settling of 10-20\% occurring for hundreds of microseconds after a stepped input signal. When appreciable settling is still occurring after hundreds of microseconds, directly digitizing the microwave reflections at GHz frequencies becomes impractical due to oscilloscope memory limitations. As discussed in section 3 and \ref{appendix:qubit_detune_spectroscopy}, we chose to use hardware demodulation to compare the performance of various packages. As a result, we are able to digitize several hundred microseconds of settling data in a single trace at the cost of some amplitude accuracy\textemdash fortunately the precision of both techniques is similar as we find relative measurements reproducible within about $\rm 2\cdot 10^{-3}$.

\begin{figure}[h]
\begin{center}
\includegraphics[width=0.75\textwidth]{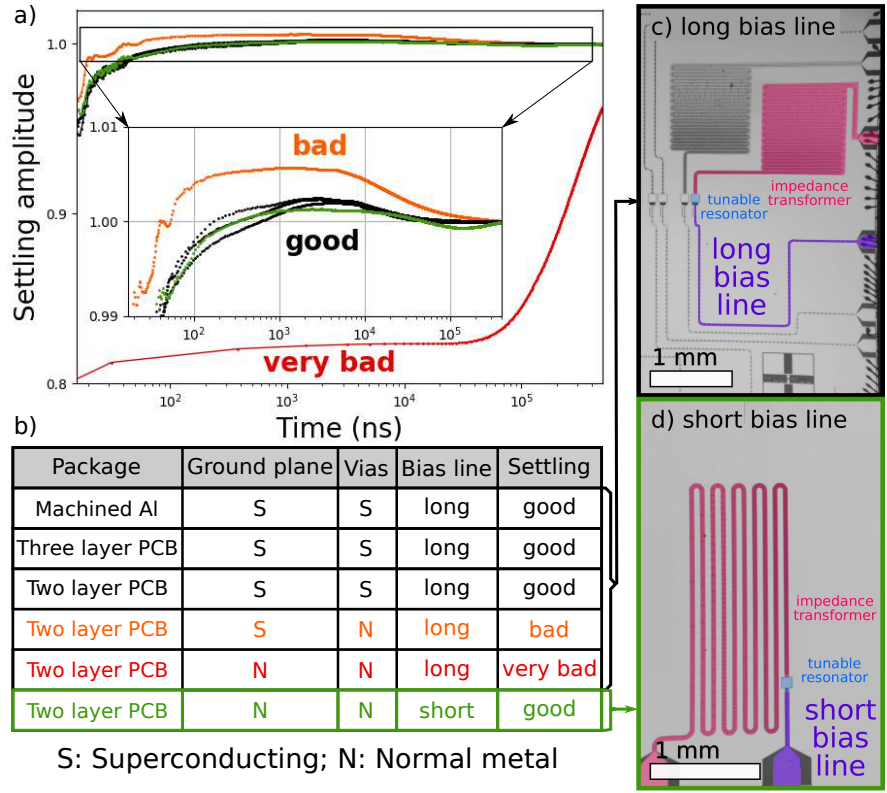}
\caption{Measured step response of various chip packages and on-chip flux bias line routings. a) Hardware demodulated step response of several different packages with two different on-chip bias line designs\textemdash the table in b) serves as a legend. b) Summary of the package and chip configurations for the traces in a). c) Micrograph image of a standard 'long' on-chip flux bias line routing. d) Micrograph image of a 'short,' symmetric on-chip flux bias line.
\label{fig:tunable_resonator_comparison}}
\end{center}
\end{figure}

In figure \ref{fig:tunable_resonator_comparison}a we plot the measured step response of a variety of packages which can be summarized as having one grouping of good settling responses (plotted together in black/green), one response that is a little bad (orange), and one that is very bad (red). The small differences between the traces in the `good' group are artifacts of hardware demodulation (see \ref{appendix:qubit_detune_spectroscopy})\textemdash the actual response of all these these packages is similar to the response in figure \ref{fig:good_data}. This grouping of four equivalently good traces includes:
\begin{enumerate}
\item Our standard machined aluminum package with normal metal signal lines
\item A two layer aluminum-plated copper PCB
\item A three layer aluminum-plated copper PCB, and finally 
\item A two layer normal metal PCB with a `short' on-chip bias line routing.
\end{enumerate}
The bad (orange) trace, which has an additional settling of $5\cdot10^{-3}$ over several hundred microseconds, is a two-layer aluminum-plated copper PCB with copper vias, and the very bad (red) trace, which has a settling of about 20\% over a ms or more is a two layer gold-plated copper PCB.

There are three conclusions to draw from these datasets. First of all, since the response of (i), (ii), and (iii) are the same, we see that the aluminum-plated copper PCBs match the fast settling response of our standard machined aluminum package, so we have a near-term scalable PCB-based packaging solution for our frequency tunable qubits. Secondly, we can compare the packaging differences between the good, bad, and very bad traces to gain insight as to the cause of the long settling time in the copper PCB. In these tests, each package is connected to normal-metal coaxial cabling on its exterior, and aluminum wire bonds connect the package to a tunable resonator chip on its interior. If we use a standard two-layer copper PCB with copper vias, we measure the very bad (red) trace. If we then plate this PCB with aluminum, but leave the vias copper, we measure the bad (orange) trace, where the settling amplitude has been reduced from 0.2 to $\rm 5\cdot 10^{-3}$. Then, if the vias are also plated in aluminum, we measure the good response (black). Again noting that the center trace of the good machined aluminum box is copper, we can conclude that these settling times are likely caused by redistribution of ground plane currents in normal metals. This current redistribution occurs as the transmission line geometry changes from coplanar waveguides on-chip, to a microstrip geometry in the package, and then to normal metal coaxial connectors as the transmission line leaves the packages. It seems plausible that some of the remaining settling in these good packages is due to the transition to the normal metal cables outside the superconducting package. 

The third and perhaps the most surprising result is that changing the on-chip bias line routing can have a significant impact on the bias line settling. The very bad (red) trace is a two-layer gold-plated copper PCB package and a typical 'long' flux bias line layout, shown highlighted purple in figure \ref{fig:tunable_resonator_comparison}c. If we use the same normal metal package with the `short' bias line layout in figure \ref{fig:tunable_resonator_comparison}d, we measure the good (green) trace in figure \ref{fig:tunable_resonator_comparison}a\textemdash indistinguishable from the long bias line layout in the good, superconducting, packages. Both the `long' and `short' bias lines have a $\rm 50\,\Omega$ coplanar waveguide geometry with superconducting aluminum crossovers every $\rm 100\,\mu m$ to connect the ground planes and short out slot-line modes. Since both bias lines have superconducting crossovers connecting the ground planes, we do not fully understand why this layout change seems to resolve the settling issue in a normal metal package. However, this indicates that on-chip layout may be able to correct for or prevent non-idealities in the package or other cabling by modifying the return current distribution. 

\section{Conclusion}
We have presented a time-domain measurement technique capable of measuring flux waveforms on a superconducting chip from DC to hundreds of MHz with a theoretical bandwidth limit of 2.6\,GHz. This transducer has a maximum flux-to-phase gain approaching $\rm 1200^\circ/\Phi_0$ that, given our phase measurement noise, translates to a flux sensitivity of $\rm 0.2\,m\Phi_0$. We have benchmarked this this technique using our standard cryogenic packaging and then applied this technique to develop a suitable PCB-based package for frequency tunable superconducting qubits. We find that for an arbitrary chip layout, packages with normal metal in the return path may introduce large settling times. This settling may be mitigated by either using superconducting aluminum-plated copper PCBs or by modifying the on-chip layout of these bias lines to avoid return current asymmetry.

\subsection*{Acknowledgments}
This work was supported by Google. Tunable resonator devices were made at the UC Santa Barbara Nanofabrication Facility, a part of the NSF funded National Nanotechnology Infrastructure Network. The PCB-based chip mount was a reproduction of a design provided by Northrop Grumman Corporation.

\section*{References}
\bibliographystyle{unsrt}
\bibliography{refs}

\newpage
\appendix
\section{Measuring microwave reflections with the tunable resonator}
\label{appendix:microwave_reflections}

At the core of the experiment described in the main text, we use the tunable resonator as a flux to microwave phase transducer. DC (or time-dependent) flux applied to the SQUID translates to a corresponding phase (or phases) of the reflected microwave tone. Ideally this allows us to sample the on-chip flux waveform and deduce the transfer function of the flux wiring. In reality, microwave reflections that occur after the probe tone has reflected off the tunable resonator will induce time-dependent phase changes which are a source of systematic error in our flux measurement. In this appendix we discuss an alternative experiment with the tunable resonator in which we use it to characterize the location and amplitude of the dominant reflections in our microwave measurement chain. This technique allows us to bound the systematic error in our flux measurement and may also be used to characterize the impedance matching of microwave components in a cryogenic environment at individual frequencies in future experiments.

\begin{figure}[h]
\begin{center}
\includegraphics[width=.8
\textwidth]{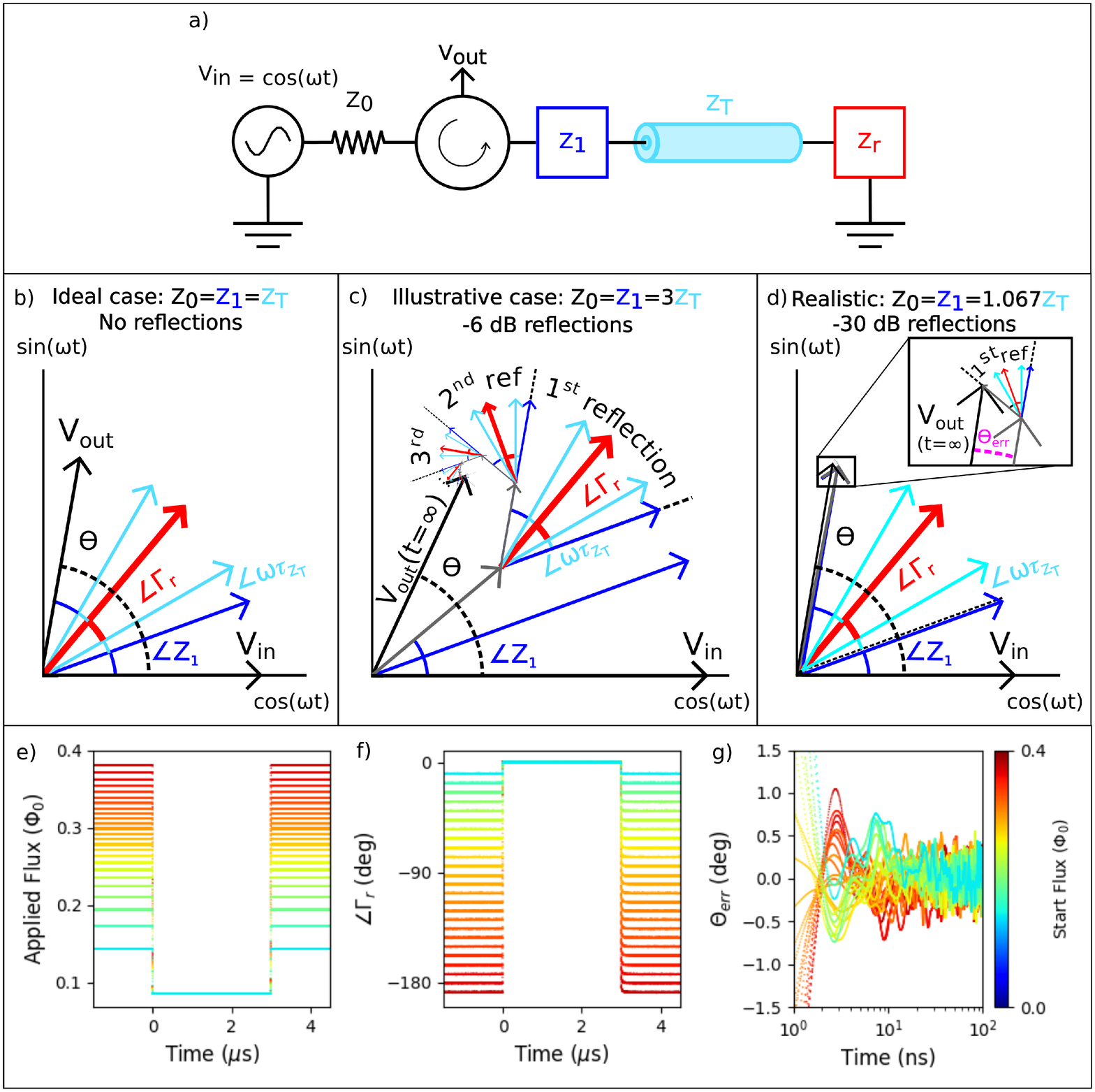}
\caption{Effect of microwave reflections on our phase measurement. a) Schematic of our phase measurement considering just two non-ideal components, $\rm Z_1$ and $\rm Z_T$, in addition to our tunable resonator $\rm Z_r$, and an ideal microwave source and circulator each with impedance $\rm Z_0$. b) Phasor diagram of the ideal case where $\rm Z_0 = Z_1 = Z_T$ and there are no unintended reflections (e.g. the only reflection is off of $\rm Z_r$). $\rm \mathbf{V_{out}}$ is just a rotation of $\rm \mathbf{V_{in}}$. c) Illustrative case considering very large, -6\,dB reflections. It takes $\rm \mathbf{V_{out}}$ many reflections to settle to a final value. d) Realistic case considering -30\,dB reflections. In this case the phasor diagram nearly collapses to the ideal case in b), but the first reflection still contributes some error $\rm \Theta_{err}$. e) Informed by the model discussed in a)-d), we use the transducer's DC flux-to-phase calibration curve to choose an array of flux waveforms that correspond to transducer reflection angles that span at least 180\,degrees as shown in f). By measuring the phase response of these steps, we vary the relative angle of the reflection and sample over at least one maxima. Finally, in g) we plot $\rm \Theta_{err}$ for 50\,ns after changing the flux. We estimate $\rm \Theta_{err}$ as the difference between each phase waveform and the average of all the phase waveforms. At 3\,ns we observe the dominant reflections where the reflection causes a spread in our phase measurement of +1.1 to -0.8 = 1.9\,degrees which corresponds to a reflection amplitude of -30 to -36\,dB.
\label{fig:reflections}}
\end{center}
\end{figure}

The schematic in figure \ref{fig:reflections}a provides a model for understanding the effect of these microwave reflections. In this model we consider two non-ideal components, $\rm Z_1$ and $\rm Z_T$, our tunable resonator $\rm Z_r$, and a microwave source and circulator each with impedance $\rm Z_0$. $\rm Z_1$ may represent any component in the actual system that causes reflections such as SMA connectors, the circulator, HEMT, \textit{etc.} and $\rm Z_T$ is a section of cabling between these components. The actual cryostat wiring consists of more than a dozen components, connections, and cables which we do not model explicitly. We instead look at the aggregate effect of all these reflections and bound their impact on our phase measurement.

We first consider the ideal case where $\rm Z_0 = Z_1 = Z_T$ and there are no unintended microwave reflections\textemdash only the reflection off of the tunable resonator $\rm Z_r$ at $\rm \angle \Gamma_r$. In figure \ref{fig:reflections}b we use a phasor diagram to plot the progression of the probe tone $\rm \mathbf{V_{in}}$ as it first leaves the microwave source, reflects off our transducer and finally propagates to our digitizer as $\rm \mathbf{V_{out}}$.  In this picture, our homodyne detector measures the magnitude and phase of $\rm \mathbf{V_{out}}$. Since there are no reflections and all of these components are assumed to be lossless, $\rm |\mathbf{V_{out}}| = |\mathbf{V_{in}}|$ and $\rm \mathbf{\hat{V}_{out}} \cdot \mathbf{\hat{V}_{in}} = \cos(\Theta)$ where $\rm \Theta$ is the sum of the accumulated phases. In our system, all the component impedances are constant except for when we change $\rm Z_r$ by applying flux. This means that for a fixed probe tone frequency, $\rm \Theta = \angle \Gamma_r + \textit{const.}$, and the constant term gets rolled into the global phase offset of the homodyne measurement leaving only the phase we are trying to measure.

If there are any mismatched impedances, microwave reflections complicate the transducer's transient response. In \ref{fig:reflections}c we consider the illustrative case of very large, -6\,dB, reflections with $\rm Z_0 = Z_1 = 3\,Z_T$. As $\rm \mathbf{V_{in}}$ leaves the generator and passes through $\rm Z_1$, it reflects off of $\rm Z_T$. This reflection effectively shifts the tunable resonator's response from the origin since half of $\rm \mathbf{V_{in}}$ reflects back to the homodyne detector without ever interacting with $\rm Z_r$. After this initial reflection, multiple reflections occur between $\rm Z_1$ and $\rm Z_r$ until the steady-state standing wave ratio is reached and $\rm \mathbf{V_{out}}$ stabalizes to $\rm \mathbf{V_{out}}(t=\infty)$. If we consider changing $\rm \angle \Gamma_r$ from one angle to another by applying a flux pulse to the tunable resonator, these reflections propagate through the system nudging the phase of $\rm \mathbf{V_{out}}$ a little bit in time every $\rm 2 \tau$ where $\tau$ is the propagation time through the transmission line $\rm Z_T$. With -6\,dB reflections, it is clear from our diagram that $\rm \mathbf{V_{out}}$ is still changing appreciably after 5 or more reflections. In figure \ref{fig:reflections}d we plot the more realistic case of -30\,dB reflections with $\rm Z_0 = Z_1 = 1.067\,Z_T$. In this case if the first reflection adds perpendicular to $\rm \mathbf{V_{out}}$, the transient phase error $\rm \Theta_{err}$ is given by $\rm \arctan(10^{-30\,dB/20\,dB}) = 1.8\,degrees$. Second reflections are negligible because a second perpendicular reflection from the same component would induce an error of $\rm \arctan(10^{-60\,dB/20\,dB}) = 0.06\,degrees$, well below the $\rm 0.25\,degree$ noise floor of our homodyne detector.

We can now apply this reflection model and use the tunable resonator transducer to estimate the amplitude and position of the dominant reflection. In the actual microwave measurement chain there are likely a number of small reflections with unknown propagation times and reflection angles. However, we have control of $\rm \angle \Gamma_r$ which is always present in the sum of angles that determine the relative angle of the reflection. If we rapidly change from some initial angle $\rm \angle \Gamma_{r,i}$ to some final angle $\rm \angle \Gamma_{r,f}$ then there will be a transient error as the first reflection propagates through the system. If we vary the starting angle $\rm \angle \Gamma_{r,i}$ over 180\, degrees while holding $\rm \angle \Gamma_{r,f}$ fixed, then regardless of the propagation time and reflection angles in the system, we will sample at least one, and maybe both of the reflection angles that maximize $\rm \Theta_{err}$. Ideally we would sample over 360 degrees, but we can only vary the transducer's reflection angle over about 200\,degrees. By sampling over 180\,degrees we can bound $\rm \Theta_{err}$ within a factor of two.

In figure \ref{fig:reflections}e we plot an array of flux waveforms that we apply to the tunable resonator to achieve the reflection angle waveforms in figure \ref{fig:reflections}f based on our transducer's DC calibration (figure \ref{fig:microwave_reflections} in the main text). The color scale in e,f and g is the same and corresponds to the initial flux value for each waveform. We choose to end all of the flux waveforms near $\rm 0.08\,\Phi_0$ where the transducer is relatively insensitive to flux\textemdash this means that the small amplitude settling from our imperfect flux transfer function will produce a negligible change to the reflection angle and allow us to distinguish reflections from flux settling. In figure \ref{fig:reflections}g we plot $\rm \Theta_{err}$ (estimated as the deviation of each $\rm \angle \Gamma_r$ waveform from the average of all the $\rm \angle \Gamma_r$ waveforms) for 50\,ns after changing the reflection angle. At $\rm t=3\,ns$ the spread in the waveforms is largest and ranges from +1.1 to -0.8 degrees and we continue to see features above the noise until about 20\,ns. This indicates that the dominate reflection in our system occurs approximately 1.5\,ns from the tunable resonator and that this reflection has an amplitude of -30 to -36 dB corresponding to $\rm \Theta_{err} = $0.85 to 1.9\,deg, depending on whether we assume $\rm \Theta_{err}$ is measured over 1 or two maximums.

In summary, reflection in the microwave side of this experiment produce time-dependent phase changes that are a source of systematic error in our transducer-based flux measurement. We use the tunable resonator to estimate the dominant reflection to be 1.5\,ns from the tunable resonator with a magnitude of -30 to -36\,dB, resulting in up to 1.8\,degrees of phase error. These reflections settle out in about 20\,ns due to the signal propagation time between the tunable resonator and the oscilloscope. This technique could also be used to characterize microwave components placed in the homodyne measurement chain anywhere from room temperature down to the cryogenic tunable resonator.

\newpage
\section{Qubit detuning spectroscopy}
\label{appendix:qubit_detune_spectroscopy}

\begin{figure}[h]
\begin{center}
\includegraphics[width=1.0\textwidth]{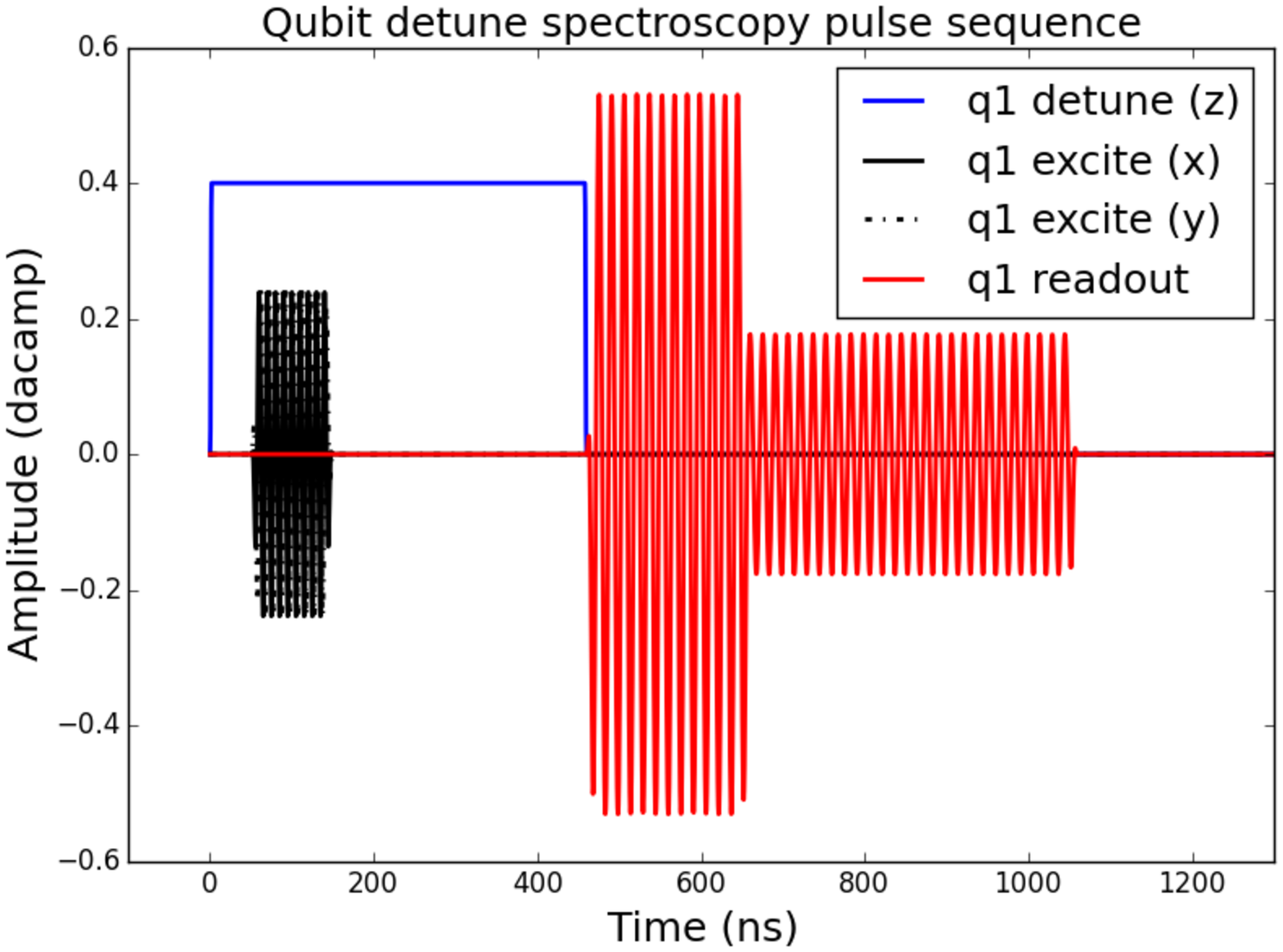}
\caption{This figure depicts a typical control pulse sequence for a qubit detuning spectroscopy measurement.  The qubit is initially prepared in the $\ket{0}$ state.  Then, at $t=0$, a flux waveform, shown in blue, is applied to the SQUID to detune the frequency of the qubit. After some variable time delay, we drive an XY pulse to try and excite qubit, then wait a fixed amount of time, detune back, and measure the state of the qubit.  By repeating this process for different delays and XY pulse frequencies, we are able to fit identify the qubit frequency in time and scale this back to an applied flux.  \label{fig:qubit_detune_pulse_sequence}}
\end{center}
\end{figure}

One way to characterize on-chip flux settling is to perform frequency detuning spectroscopy with qubits\textemdash figure \ref{fig:qubit_detune_pulse_sequence} depicts a typical pulse sequence. The qubit is first prepared in the ground state. At time $\rm t=0$ a frequency detuning flux pulse is applied to change the qubit frequency. After a variable amount of time we drive the qubit with a variable frequency XY pulse to try and excite the qubit before detuning back to the qubit readout frequency for measurement. By varying the delay time between the start of the flux pulse and the XY pulse and frequency of the XY pulse, we are able to measure the qubit frequency in time and map that back to an applied flux.

\begin{figure}[h]
\begin{center}
\includegraphics[width=1.0\textwidth]{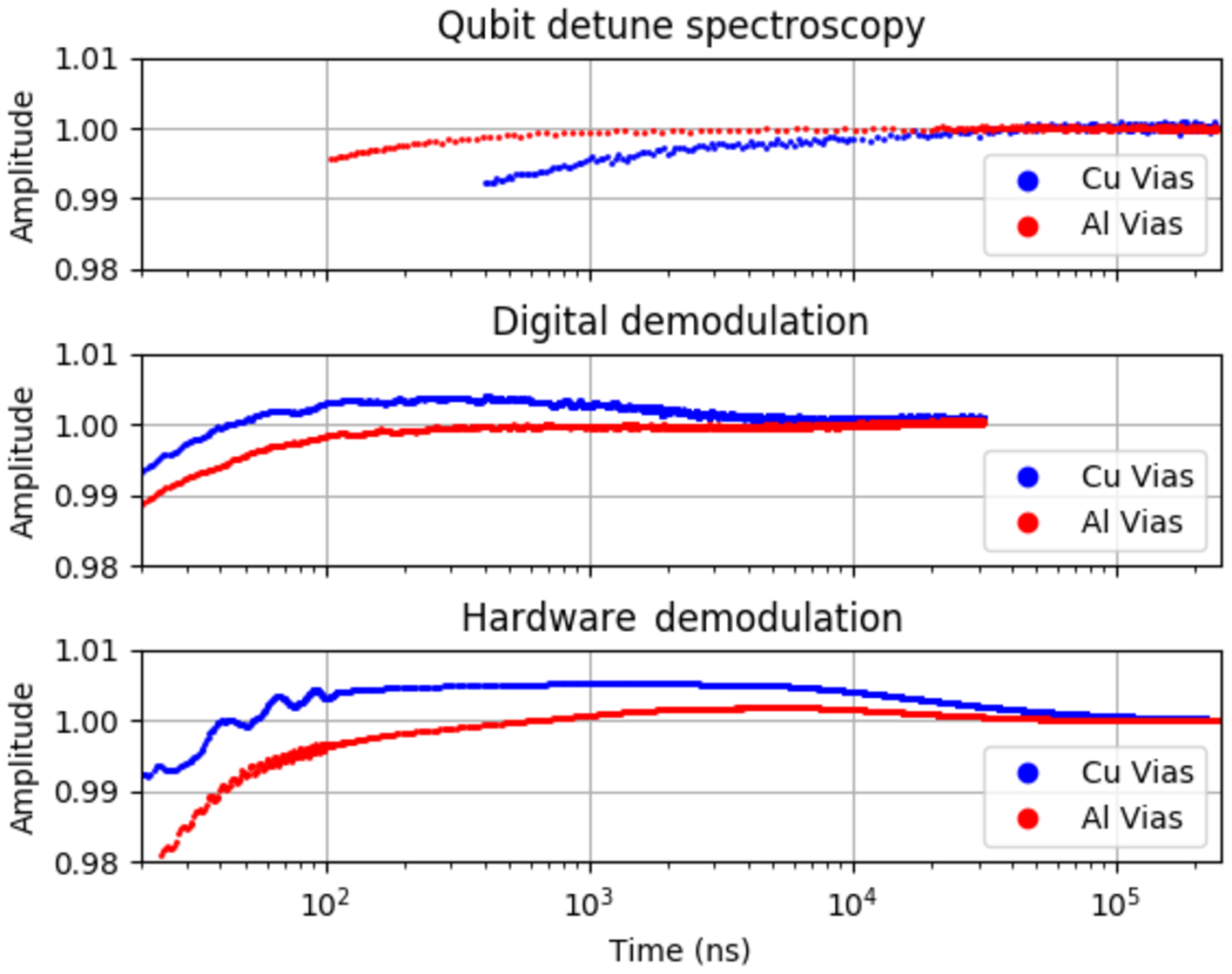}
\caption{Measurement of flux bias line settling comparing qubit detuning spectroscopy to hardware and digitally demodulated data from the tunable resonator. We plot the measured response of both the three layer aluminum PCB with aluminum vias (red) and the two layer PCB with copper vias (blue). In each case, we are able to reproduce very small additional settling amplitude of about 0.005 over 10's of $\mu s$. While the use of hardware demodulation allows us to measure at longer times, sampling a DC step response subjects us to the DC settling response of our oscilloscope which imparts a settling of 0.002 over 10's of $\mu s$ even with the good aluminum via package. The fact that the aluminum via package has a very flat response in both the qubit data and digitally demodulated tunable resonator data indicates that our AWG has a very flat output. Finally, one may note that the copper via device undershoots in the qubit data but overshoots with the tunable resonator data (middle and bottom).  This is likely due to the fact that the on-chip layout of these lines was not identical between the qubit and tunable resonator devices.
 \label{fig:qubit_vs_dig_vs_mixer}}
\end{center}
\end{figure}

In figure \ref{fig:qubit_vs_dig_vs_mixer} we plot the measured step response of the nominally good three-layer aluminum PCB with aluminum vias and the two-layer aluminum PCB with copper vias that exhibits a settling amplitude of about $\rm 5\cdot 10^{-3}$ over $\rm 10-100\,\mu s$. The key result is that regardless of the measurement technique used, we are able to reproduce a similar difference in settling behavior between these two PCBs despite it being less than 1\%. The next observation is that the good aluminum via package (red) has a very flat settling in both the qubit and digital demodulation datasets where-as the hardware demodulated data shows a settling behavior of $\rm 2\cdot 10^{-3}$ over 10's of $\rm \mu s$ even on the aluminum via package. While using hardware demodulation allows us to sample at a lower rate and measure at longer times (accommodating the limited oscilloscope memory), we are subjected to the oscilloscope's impulse response, which is not the case for both the qubit and digitally demodulated datasets.  The fact that both the qubit and digitally demodulated data have a very flat response indicates that our AWG performs well with respect to settling.  Finally, notice that the copper via data under-shoots in the qubit data and overshoots in the tunable resonator data for both digital and hardware demodulation.  This is likely because the exact on-chip bias line routing differences between the qubit data and the tunable resonator data; while we have not studied this carefully, we have seen evidence across multiple devices in different packages that certain on-chip routings overshoot while others undershoot. This is not necessarily surprising given the stark difference in settling between the long and short flux bias lines discussed in figure \ref{fig:tunable_resonator_comparison} in the main text.

\end{document}